\documentclass{emulateapj}

\shortauthors{Shi et al.}

\begin{document}

\title{Unobscured Type 2 AGNs}

\author{Yong Shi\altaffilmark{1}, George H. Rieke\altaffilmark{2}, Paul Smith\altaffilmark{2}, Jane Rigby\altaffilmark{3}, 
Dean Hines\altaffilmark{4}, Jennifer Donley\altaffilmark{5}, Gary Schmidt\altaffilmark{6},  Aleksandar M. Diamond-Stanic\altaffilmark{2}}

\altaffiltext{1}{Steward Observatory, University of Arizona, 933 N Cherry Ave, Tucson, AZ 85721, USA; currently at Infrared Processing and Analysis Center, Pasadena, CA}
\altaffiltext{2}{Steward Observatory, University of Arizona, 933 N Cherry Ave, Tucson, AZ 85721, USA}
\altaffiltext{3}{Observatories, Carnegie Institution of Washington, 813 Santa Barbara St., Pasadena, CA 91101, USA ; Spitzer Fellow}
\altaffiltext{4}{Space Science Institute, Boulder, CO.}
\altaffiltext{5}{Steward Observatory, University of Arizona; currently at Space Telescope Science Institute, Baltimore, MD}
\altaffiltext{6}{Steward Observatory, University of Arizona; currently at National Science Foundation, Arlington, VA}

\begin{abstract}

Type 2 AGNs with intrinsically  weak broad emission lines (BELs) would
be  exceptions to  the  unified  model. After  examining  a number  of
proposed   candidates  critically,   we  find   that  the   sample  is
contaminated   significantly  by  objects   with  BELs   of  strengths
indicating that  they actually contain intermediate-type  AGNs, plus a
few Compton-thick sources as revealed by extremely low ratios of X-ray
to nuclear IR luminosities.  We develop quantitative metrics that show
two (NGC 3147  and NGC 4594) of the remaining  candidates to have BELs
2-3  orders   of  magnitude  weaker  than  those   of  typical  type-1
AGNs. Several  more galaxies remain as candidates  to have anomalously
weak  BELs, but  this status  cannot  be confirmed  with the  existing
information.  Although  the parent sample  is poorly defined,  the two
confirmed objects are  well under 1\% of its  total number of members,
showing that  the absence of a  BEL is possible, but  very uncommon in
AGN. We  evaluate these two  objects in detail  using multi-wavelength
measurements  including new IR  data obtained  with {\it  Spitzer} and
ground-based  optical  spectropolarimeteric  observations.  They  have
little   X-ray  extinction  with   $N_{\rm  H}$   $<$  $\sim$10$^{21}$
cm$^{-2}$.  Their IR spectra  show strong silicate emission (NGC 4594)
or weak  aromatic features on a  generally power law  continuum with a
suggestion of silicates  in emission (NGC 3147).  No  polarized BEL is
detected in NGC  3147. These results indicate that  the two unobscured
type-2 objects have circumnuclear tori that are approximately face-on.
Combined  with their  X-ray and  optical/UV properties,  this behavior
implies that  we have an unobscured  view of the nuclei  and thus that
they have  {\it intrinsically} weak BELs. We  compare their properties
with those of  the other less-extreme candidates. We  then compare the
distributions of bolometric luminosities  and accretion rates of these
objects with theoretical models that predict weak BELs.

\end{abstract}

\keywords{galaxies: Seyfert -- galaxies: nuclei}

\section{Introduction} 
 
Active galactic nuclei (AGNs) are optically classified into type 1 and
type 2:  a type 1  object shows broad  emission lines (BELs)  that are
absent in  type 2s.  The strict AGN  unified model \citep{Antonucci93,
  Urry00} states  that the apparent large diversity  in AGN properties
is   caused   almost entirely  by   different   viewing   angles  and   nuclear
luminosities. An  AGN is identified  as type 2  when a dusty  torus is
viewed  at  large inclination  (more  edge-on)  and  obscures the  BEL
region, while a  torus in a type 1 AGN is  viewed at lower inclination
and does not block the BEL region.

In  the  past few  decades,  the  unified model  has  achieved
tremendous  success  in explaining  the  general  properties of  AGNs.
However, there are  indications that its axioms may  not be universal;
parameters  other  than orientation  may  influence  the observed  AGN
properties.  For example,  the host galaxy may play  an important role
in the type  1/type 2 division.  It has been  found that a significant
fraction of type 2 AGNs are actually obscured by kpc-scale material in
the    host   galaxy    instead   of    the   sub-kpc    dusty   torus
\citep[e.g.][]{Keel80, Maiolino95a,  Rigby06, Diamond-Stanic09a}.  The
typical level of star formation in  a type 2 host galaxy is also found
to be  higher than that in  a type 1  host \citep{Maiolino95b, Shi07},
implying that  a more dusty type  2 host increases  the probability of
obscuration  of   the  BEL  region.   In   radio-loud  AGNs,  infrared
observations  indicate that  a significant  fraction of  type  2 radio
galaxies  harbor  intrinsically weaker  nuclei  than  type 1  galaxies
\citep{Whysong04,  Shi05, Shi07,  Ogle06, Cleary07}.   There  are also
claims  that  AGNs with  intrinsically  weak  BEL  emission may  exist
\citep[e.g.][]{Laor03, Hawkins04,  Diamond-Stanic09b, Hopkins09}.  For
example,  spectropolarimetric observations  reveal  hidden BEL  (HBEL)
regions  in  only  half  of  type 2  Seyferts  \citep{Kay94,  Moran00,
  Tran01}.  The comparison between Seyfert 2 galaxies with and without
HBELs  reveals  a significant  difference  in  the nuclear  luminosity
between  the two  populations  \citep{Tran01, Gu02}.   However, it  is
unclear  if the  low nuclear  luminosity of  Seyfert 2  nuclei without
HBELs is simply due to a  larger host galaxy contamination for a given
nuclear  luminosity  or if  Seyfert  2  galaxies  without HBELs  harbor
genuinely weak nuclei.

In  the standard  model, a  type 2  object shows  only  narrow optical
emission lines in  total light and has strong  X-ray obscuration.  The
HI  equivalent   column  density  can  be   inferred  through  fitting
photoelectric absorption  models to the X-ray  spectrum.  As expected,
the  HI column  densities  of 96\%  of  type 2  AGNs  are larger  than
10$^{22}$  cm$^{-2}$, and they  are therefore  considered to  be X-ray
obscured \citep{Risaliti99}.   However, with the  increasing volume of
X-ray spectral data,  a small sample of type 2 AGNs  has been found to
be   relatively  unobscured  in   the  X-ray   \citep{Pappa01,  Xia02,
  Panessa02,  Georgantopoulos03a,   Wolter05,  Gliozzi07,  Bianchi08a,
  Brightman08}.  In  the literature, they  are defined as type  2 AGNs
with $N_{\rm H}$ $<$ 10$^{22}$ cm$^{-2}$.  This sample of unusual AGNs
may offer the best means to  refine the unified model and may identify
exceptions  to the model  that would  modify our  overall view  of the
subject.

Various explanations  have been proposed  to understand the  nature of
this  unusual  type of  AGN:  (1)  Compton-thick  behavior: The  X-ray
spectra of  Compton-thick objects in the  energy range $<$  10 keV are
dominated by the  reflection of the nuclear emission  off the far side
of the torus and the  HI column densities inferred from X-ray spectral
fits are  similar to those  for unobscured objects; (2)  Dilution: The
BELs  in these  objects are  overwhelmed  by host  galaxy light.   (3)
Variability:  Some  AGNs  change  their optical  classification  on  a
timescale of years to decades as a result of the location of obscuring
material on pc scales.  Such objects might be classified as unobscured
(in the X-ray) type 2 objects by chance, if the X-ray and optical data
are  obtained at  different times.   (4) S/N  effect: The  BEL  can be
hidden in a low-S/N optical spectrum while the Compton-thick signature
can be  lost in a  low S/N X-ray  spectrum; (5) Exceptions to  the AGN
unified model:  in contrast to  the above speculations that  still fit
the  unified model,  these  X-ray-unobscured type  2  AGNs may  harbor
genuinely  weak BEL  regions, and  thus  are not  the simple  obscured
version of type 1 AGNs expected from the unified model.

In this paper,  we will first combine {\it Spitzer}  IR data with data
in the literature to update  the current list of X-ray-unobscured type
2 AGN candidates.  We then review the members of this list to identify
objects  that can be  explained by  points 1  - 4  above and  that are
therefore  not  true  unobscured  type-2 objects. Of 24 candidates from
the literature, less than a third are possible true unobscured  type-2 AGNs. We
evaluate these candidates, including using metrics for relative broad line strength, and find that the broad
lines in two of them are at least two orders of magnitude weaker than 
in typical type-1 AGNs. Their IRS/Spitzer spectra indicate that 
we have  a pole-on view  of their circumnuclear tori,  consistent with
the  X-ray  measurements  of  small  absorbing  columns  toward  their
nuclei.  Thus,  the  properties  of  these two  AGNs  appear  to  be
contradictory  between type-1  levels of  obscuration but  type-2 line
properties. We conclude that AGN with intrinsically weak BELs are
very rare, but that they do exist.    

We   develop   our    arguments   as    follows.    In
Section~\ref{DATA},  we   describe  the  sample   selection  and  data
analysis.  In  Section~\ref{CONT_UNAGN}, we eliminate  contaminants to
the sample of X-ray-unobscured type  2 AGNs, including type 1 AGNs and
Compton-thick objects.   The IR observations of  the remaining objects
are presented  in Section~\ref{discussion} and placed  in context with
observations  at other wavelengths.   Section~\ref{conclusion}
summarizes our conclusions.

\section{Sample Selection And Data Reduction}\label{DATA}

\subsection{Sample Selection}

Our initial sample is listed in Table~\ref{sample_opt} and is composed
of 24 objects identified as  X-ray-unobscured type 2 AGN candidates in
the literature  (``candidate'' refers to an un-verified suggestion of an X-ray-unobscured
type 2 AGN).  \citet{Panessa02} compiled  a total of 17
candidates. However, as listed in  their Table 2, NGC 6221 has $N_{\rm
  H}$ $>$ 10$^{22}$ cm$^{-2}$ and is thus excluded from our list.  NGC
2992  is   also  not  included   as  it  has  variable   $N_{\rm  H}$,
1.4$^{+0.5}_{-0.4}$$\times$10$^{22}$    cm$^{-2}$    in    1997    and
0.90$\pm$0.03$\times$10$^{22}$ cm$^{-2}$  in 1998 \citep{Gilli00}.  IC
1631 is also excluded, since the XMM-Newton observation indicates that
it  has  an  X-ray flux  three  orders  of  magnitude lower  than  the
upper-limit  based  on  the  GINGA  measurements  originally  used  in
\citet{Panessa02},   and  it  is   thus   most    likely   Compton-thick
\citep{Bianchi08b}.

Among objects with AGN types $>$ 1.5 and X-ray columns or upper-limits
$<$  10$^{22}$ cm$^{-2}$ in  a complete  Seyfert sample  observed with
{\it XMM-NEWTON} by \citet{Cappi06}, we identified two more candidates
(NGC 3941 and NGC 4501).  Confirmed and possible Compton-thick objects
(see their Table 4) are not included.  However, the new observation of
NGC 4501 with Chandra indicates  that the observed low HI column density with
XMM-Newton is mainly caused by significant extra-nuclear contamination
\citep{Brightman08}.  NGC  4501 is most likely  a Compton-thick object
and thus excluded from our sample.

\subsection{Optical Emission Line Data}

These objects were originally classified by different authors. We have
re-classified them in a consistent way using the classification scheme
in \citet{Ho97a}. The result  is listed in Table~\ref{sample_opt}.  For
each   object,  we   have  compiled from  the   literature  the
[OIII]$\lambda$5007  flux   and  the  narrow   line  ratios  including
$\frac{\rm [SII]}{\rm  H{\alpha}}$, $\frac{\rm [OIII]}{\rm H{\beta}}$,
$\frac{\rm    [OI]}{\rm    H\alpha}$    and   $\frac{\rm    [SII]}{\rm
  H{\alpha}}$. The extinction correction for the line emission follows
that of \citet{Ho97a}.

\subsection{Optical Spectropolarimeteric Data}

To search  for possible HBELs in  reflection, new spectropolarimeteric
observations were obtained for NGC  3147, NGC 4698 and RXJ 1737.0+6601
with  the 2.3 m  Bok Reflector  on Kitt  Peak.  The  observations were
carried  out on March  30-31, 2008,  using the  CCD Spectropolarimeter
\citep[SPOL;][]{Schmidt92},  upgraded  with  a 1200$\times$800  pixel,
thinned,  antireflection-coated, UV-sensitized  CCD,  and an  improved
camera lens and half-wave plate.  SPOL was configured using a 600 l/mm
grating  that  covers  4400-8200  $\AA$  so that  both  H$\alpha$  and
H$\beta$  were  included  in  the  spectrum  for  all  three  objects.
Combined with a slit width of 2$''$, the resolution was $\sim$15 $\AA$
(900 and  700 km/s at H$\beta$ and  H$\alpha$, respectively), adequate
to identify BELs (FWHM $>$ 1000 km/s).  A Y-46 filter was also used to
block second-order contamination  at H$\alpha$ for RXJ1737.0+6601. The
observations were made on dark  nights to minimize the sky signal.
The total integration  time was 12,800 sec for NGC  3147 and NGC 4698,
and  15,360  sec  for  RXJ1737.0+6601.   The  data  reduction  follows
\citet{Smith03} and the results are shown in Table~\ref{POL_RESULT}.

\subsection{X-ray Data}

X-ray data,  including observatory, X-ray flux, HI  column density and
the equivalent width (EW) of the Fe K$\alpha$ line, were compiled from the
literature and  listed in Table~\ref{sample_X-ray}.  Whenever multiple
observations were available, all are included to evaluate the X-ray
variability  and the  dependence  of  the X-ray  flux  on the  spatial
resolution.

\subsection{Infrared Data}\label{IR-DATA}

We obtained the {\it  Spitzer} IRAC photometric and IRS low-resolution
spectroscopic data  in our own  program (PID-40385, PI-G.   Rieke) and
from  other  archived programs,  as  listed in  Table~\ref{sample_IR}.
The IRAC photometry was measured  from the post-BCD
images.  For the IRS mapping mode, all BCD images were sky-subtracted,
bad-pixel-masked  and combined to  build the  final cube  using CUBISM
\citep{Smith07}.  For the IRS staring  mode, the BCD image was cleaned
with IRSCLEAN and sky-subtracted using the BCD images of another order
but the same module  \citep[For a detailed description, see][]{Shi09}.
The archived 2MASS images at $J$, $H$ and $K$ were also retrieved.

Except for the  quasars, the objects were well-resolved  with the {\it
  Spitzer} beam. To  construct  the IR  spectral  energy  distribution  (SED), we  have
adopted different strategies for: (1) quasars; (2) objects mapped with
IRS; and (3)  nearby objects observed only with  the IRS staring mode.
For quasars,  the SED of the  total IR emission  was constructed.  The
IRAC photometry was  measured using a 5 pixel  radius with sky annulus
between 5  and 10 pixels (1.2$''$/pixel).   The corresponding aperture
corrections were 1.061, 1.064, 1.067  and 1.089 for the IRAC 3.6, 4.5,
5.8  and 8.0  $\mu$m  channels, respectively.   The  IRS spectrum  was
extracted  using the  optimal  extraction algorithm  with SPICE.   The
2MASS  photometry  was  extracted  using  an  aperture  of  10  pixels
(1$''$/pixel) and  a sky annulus of  [10, 15] pixels  and the aperture
correction was obtained using a star image.

For  nearby   objects  observed  with   the  IRS  mapping   mode  (see
Table~\ref{sample_IR}),  the  extraction  aperture  for  all  IR  data
(2MASS, IRAC and IRS) was defined using the IRAC-5.8$\mu$m image.  The
emission at  5.8 $\mu$m  is dominated by  hot dust radiation  and best
locates  the active  nuclei.  The  aperture was  centered at  the peak
emission  of the  5.8 $\mu$m  image  and had  a diameter  of 9  pixels
(equivalent to the slit width  of the IRS long-low module).  The 2MASS
and IRAC  photometry were  measured within this  aperture and  the IRS
spectrum  was also  extracted using  CUBISM.  No  aperture corrections
were applied  for these  extended objects.  Slight  mismatches between
different  spectral   orders,  due  to   astrometric  and  photometric
uncertainties, were removed by  re-scaling the longer wavelength order
to the  next lower  one.  The scaling  adjustments were  $<$10\%.  The
final IRS spectra were then  re-normalized to the IRAC photometry at 8
$\mu$m requiring adjustments by $<$30\%.

For nearby  objects observed with  the IRS staring mode,  the spectra
were extracted using the  regular extraction algorithm with SPICE.  The
slit  widths of the  IRS short-low  (5.2-14.5$\mu$m) and  IRS long-low
(14.0-38.0$\mu$m) modules are  3.7$''$ and 10.7$''$, respectively. The
aperture for the IRAC and  2MASS photometry was defined to be centered
at the  IRS extraction aperture and  had a size of  7$''$ in diameter,
which gave the minimum difference  of $<$20\% between the IRAC and IRS
flux densities  at IRAC 8  $\mu$m. No aperture correction  was applied
for these extended objects.

\begin{figure*}
\epsscale{.90}
\plotone{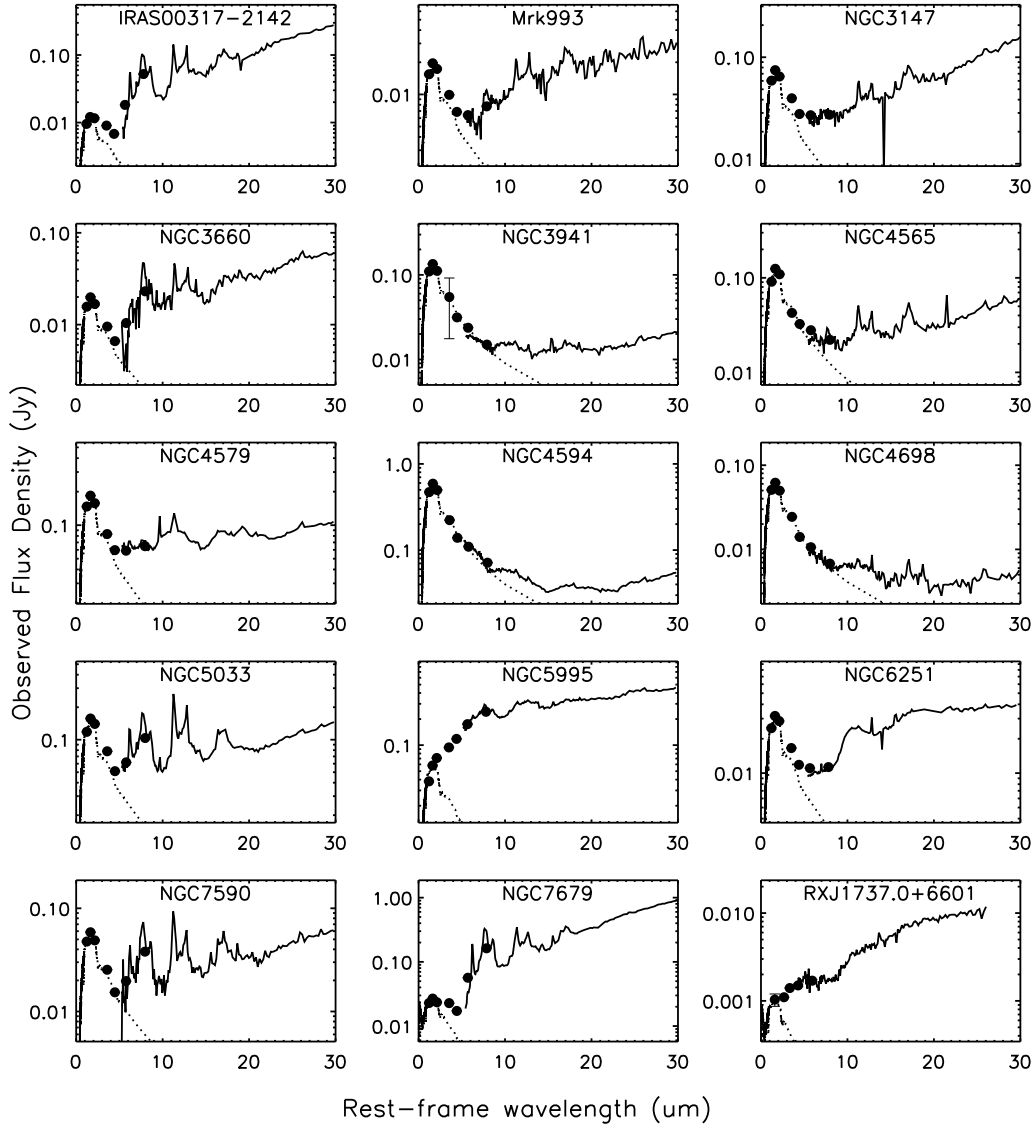}
\caption{\label{OBS_SPEC} The observed IRS spectra (solid lines) 
and broad-band photometry (circles) including 2MASS and IRAC bands.
The dotted line is the best \citet{Bruzual03} model fitted to 2MASS photometry. }
\end{figure*}

The IRS spectra are  shown in Figure~\ref{OBS_SPEC}.  For each object,
the spectrum  of the stellar photospheric emission  was estimated from
stellar models.   To determine the  appropriate model, we  extracted a
series of  single stellar population  (SSP) models with  ages spanning
from  0.01 Gyr  to 12  Gyr and solar  metallicity \citep{Bruzual03}.  The adopted model provides the
best match to the 2MASS photometry by minimizing the $\chi^{2}$ value,
shown  as dotted  lines.  For  the  quasar RXJ  1737.0+6601 with  only
K-band  photometry, the oldest  stellar model  was adopted.   For each
object, we  calculated the nuclear  IR emission from the  dusty torus.
The flux  densities in the IRAS  12 $\mu$m and MIPS  24 $\mu$m filters
were measured  with  the stellar-subtracted  IRS spectra.   The
star-forming component was further subtracted based on the 11.3 $\mu$m
aromatic flux scaled to  the \citet{Rieke09} template.  The result for
the nuclear IR flux density is listed in Table~\ref{sample_IR}.

\section{Identification of Unobscured Type 2 AGNs}\label{CONT_UNAGN}

\subsection{Contaminants}

Many  sources are  incorrectly identified  as X-ray-unobscured  type 2
AGNs  \citep[e.g., ][]{Lumsden04}.  These  objects must  be  classified
carefully and rejected from the  sample. We list the AGN
that     we    eliminated     from    our     initial     sample    in
Table~\ref{mis_classification} and discuss them below.

\subsubsection{AGNs With BELs}\label{MISS_AGNTYPE}

\begin{figure}
\epsscale{.90} 
\plotone{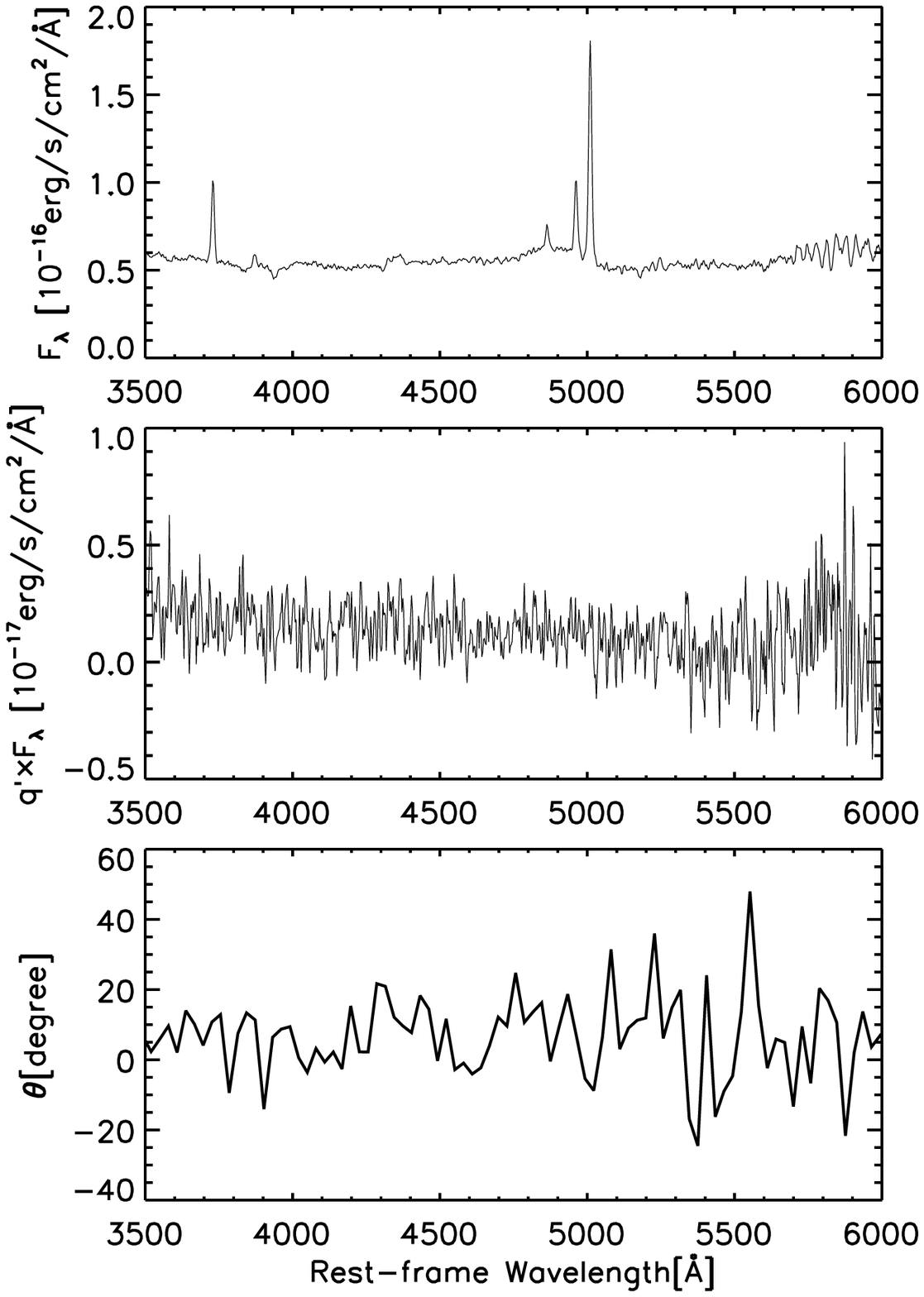}
\caption{\label{RXJ1737_SPEC} 
The  presence  of  a  broad  H$\beta$  line in  the  spectrum  of  RXJ
1737.0+6601   previously   classified   as   a  type   2   quasar   in
\citet{Wolter05}, as indicated in  the upper panel. The polarized flux
density is  shown in  the middle panel.   The linear  Stokes parameter
q$'$ has been rotated so  that the polarization position angle is zero
averaged over  the entire spectrum.  The increased  noise at $\lambda$
$>$  5700$\AA$ is  due  to the  appearance  of terrestrial  atmosphere
features  in  the near-IR.  The  lower  panel  shows the  polarization
position   angle   where  the   wavelength   has   been  smoothed   to
$\Delta\lambda$ of 40$\AA$.}
\end{figure}

As shown  in Table~\ref{sample_opt}, we have  reclassified all objects
in our  initial sample  on a consistent  basis as either  HII galaxies
(dominated by  star formation),  LINERs, Seyfert galaxies,  or quasars
($L_{\rm  X}$ $>$  10$^{44}$ erg  s$^{-1}$), following  the  scheme of
\citet{Ho97a}. This  approach is based on narrow  emission line ratios
including  $\frac{\rm [SII]}{\rm  H{\alpha}}$,  $\frac{\rm [OIII]}{\rm
  H{\beta}}$, $\frac{\rm [OI]}{\rm H\alpha}$ and $\frac{\rm [SII]}{\rm
  H{\alpha}}$.  The presence and strength of a BEL was further used to
assign intermediate types (1.2, 1.5, 1.8 \& 1.9) to objects classified
as AGNs  (LINERs, Seyferts and  quasars).  Two objects  have ambiguous
classifications.   NGC 4579  is at  the edge  of the  division between
Seyfert and  LINER regimes.   NGC 5995 is  classified as  a HII-galaxy
based  on  $\frac{\rm  [OIII]}{\rm  H{\beta}}$$-$$\frac{\rm  [OI]}{\rm
  H\alpha}$   and   $\frac{\rm   [OIII]}{\rm   H{\beta}}$$-$$\frac{\rm
  [SII]}{\rm  H{\alpha}}$   diagnostics,  but  has   a  solid  Seyfert
classification     based     on     the     $\frac{\rm     [OIII]}{\rm
  H{\beta}}$$-$$\frac{\rm   [NII]}{\rm   H{\alpha}}$.  Its IRS spectrum 
(Figure~\ref{OBS_SPEC}) is typical of a moderately obscured AGN. The   optical
spectrum of  Mrk 273x does  not cover the  H$\alpha$ line and  its AGN
nature was deduced from its high X-ray luminosity ($\sim$10$^{44}$ erg
s$^{-1}$) \citep{Xia99}. XBSJ  031146.1-550702 does not have published
line   ratios   and   the    S1.9   classification   is   taken   from
\citet{Caccianiga04}.   NGC  6251   is   classified  as   type  2   in
\citet{Shuder81} but shows  a BEL in very narrow-slit  {\it HST} spectra
\citep{Ferrarese99, VerdoesKleijn06}.  From the ratio of broad to narrow
line strengths in \citet{Ferrarese99}, it would appear to be 
an intermediate-type; it could also be variable in type. The strong
silicate emission in its spectrum (Figure~\ref{OBS_SPEC}) 
supports a type-1 designation. For one of
the  three   type  2  quasars  identified   in  \citet{Wolter05}  (RXJ
1737.0+6601), our  new high-S/N observation reveals  a prominent broad
H$\beta$ line as shown in Figure~\ref{RXJ1737_SPEC}, identifying it as
a type  1 quasar.  The  power-law-dominated IR spectrum 
(Figure~\ref{OBS_SPEC}) is consistent
with  this classification. The
remaining  two  objects (RXJ  1715.4+6239  and  RXJ 1724.9+6636)  will
require high S/N optical spectra to classify them reliably.

\begin{figure}
\epsscale{1.0}
\plotone{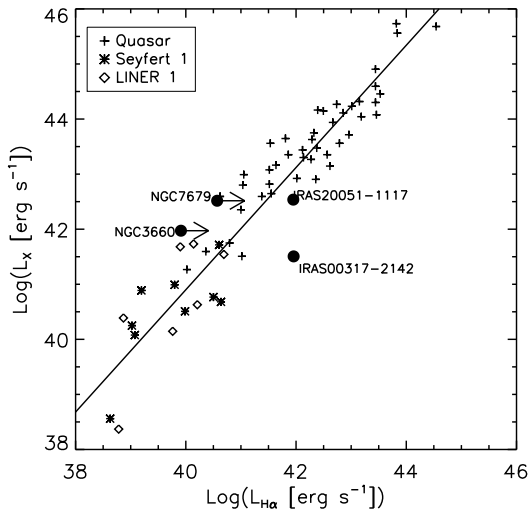}
\caption{\label{Lx_Lha} The correlation between the X-ray luminosity
  $L_{X}$ and the  H$_{\alpha}$ luminosity $L_{\rm H_{\alpha}}$  for
  quasar (plus sign), Seyfert 1 (asterisk) and LINER 1 (diamond) from \citet{Ho01}, where the
  extinction-corrected $L_{X}$ and the total H$\alpha$ luminosity
  corrected for  extinction derived from line ratios measured for the narrow line regions are plotted. 
  The  solid line is  the linear fit  to these objects. 
  Four objects classified as HII regions based on optical line ratios are 
  shown as filled circles; the broad H$\alpha$ luminosities uncorrected
  for reddening are plotted as lower-limits for NGC 3660 and NGC 7679. The figure shows
  that even  the observed broad  H$\alpha$ luminosities  of two  objects are  not  significantly lower   than  the  total
  unabsorbed   H$\alpha$   luminosities    expected   from   the   X-ray
  luminosities. }
\end{figure}

The   initial   sample   of    X-ray-unobscured   type   2   AGNs   in
Table~\ref{sample_opt}  is  contaminated by  four  HII galaxies  (IRAS
00317-2142, NGC  3660, IRAS 20051-1117  \& NGC 7679) whose  IR spectra
are dominated  by aromatic features.  All  of them show  BELs in their
optical  spectra.  They  were  misclassified  as type  2  AGNs in  the
literature   most   likely   because   intense   emission   from   the
star-formation   region  dilutes   the  BEL   emission  significantly.
Figure~\ref{Lx_Lha}     shows     the     relation     between     the
extinction-corrected   X-ray  luminosity   and  the   total  H$\alpha$
luminosity corrected for narrow-line extinction \citep{Ho01}.  Two HII
objects  (IRAS 00317-2142  and IRAS  20051-1117) are  also  plotted in
Figure~\ref{Lx_Lha}. For  the remaining two  objects (NGC 3660  \& NGC
7679), the H$\alpha$ luminosities  uncorrected for reddening are shown
as lower-limits.   The broad component  of the H$\alpha$ line  for NGC
3660 is estimated from  the published spectrum in \citet{Goncalves99}.
The X-ray luminosities  are from Chandra data for  IRAS 00317-2142 and
IRS 20051-1117.   As shown  in Figure~\ref{Lx_Lha}, even  the observed
broad H$\alpha$ luminosities of two objects (NGC 3660 \& NGC 7679) are
not   significantly  lower   than  the   total   unabsorbed  H$\alpha$
luminosities  expected  from  the  X-ray  luminosities.   This  result
implies that the AGNs in  these HII galaxies do not have intrinsically
weak BELs and thus should be excluded from the sample.

For the 20 objects classified  as AGNs (LINERs, Seyferts, or quasars),
only seven  are of type 2.   To be conservative, we  have excluded Mrk
273x,  since  its optical  spectrum  does  not  cover H$\alpha$.   The
remaining objects  show BELs at different  levels with classifications
ranging from Type 1.2 to type 1.9.  Note that the Compton-thin L1.9/S1.9
galaxy NGC  4579 shows a strong silicate  emission feature, indicating
its  weak narrow  H$\alpha$  emission  is not  likely  to result  from
extinction.   However,  the  detection  of an  additional  very  broad
H$\alpha$ component indicates that the  total emission in this line is
significantly   larger   than    in   the   narrow   component   alone
\citep{Barth01}.

\subsubsection{The Polarization of RXJ 1737.0+6601}

Significant linear  polarization was discovered in the optical continuum of RXJ
1737.0+6601. The narrow and  broad lines appear
to   be  unpolarized   (Figure~\ref{RXJ1737_SPEC})  at   a  1-$\sigma$
polarization level of 1\%.  If  the polarization is due to scattering,
the scattering material needs to be relatively close to the BEL region
in order to account for  the lack of line polarization. Alternatively,
the  featureless polarized  spectrum could  result from  a synchrotron
continuum.   Variability  of  the  polarization or  the  detection  of
scattered H$\beta$  in future high-S/N spectro-polarimetry  of RXJ 1737.0+6601
would be definitive in identifying  the source of polarized flux. 

\subsubsection{Compton-thick AGNs}\label{COMPTON_AGNs}

\begin{figure}
\epsscale{.90}
\plotone{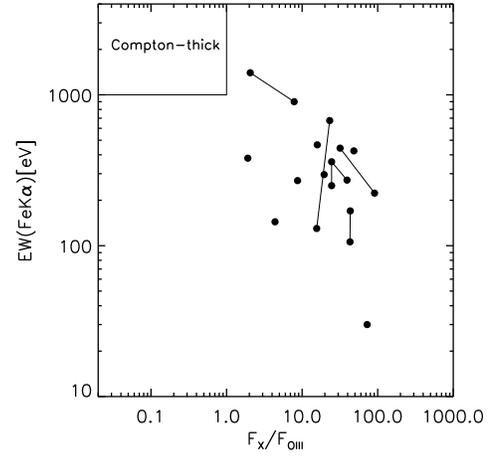}
\caption{\label{T_EW_FeKa} Distribution of X-ray unobscured type 2 AGN candidates in equivalent 
width of Fe K$\alpha$ versus the ratio of X-ray and [OIII] 
line fluxes. Pairs of data points connected with lines represent 
multiple X-ray observations for the same object. The figure shows that
these objects  have $F_{X}$/[OIII] = $T$  $\ge$ 1  and EW(FeK$\alpha$)  $\lesssim$ 1  keV, i.e.,
outside  the  locus  of  the  Compton-thick  region.  }
\end{figure}

In Compton-thick AGNs, the transmitted photons are completely absorbed
at  2 -  10 keV and the emission is  dominated by
reflected emission  from cold or warm  scatterers \citep{Matt00}.  The
objects may be thus  mis-classified as having low absorption. However,
they can still be identified  through the much lower luminosity of the
reflected component  and, usually, by the large  equivalent width (EW)
of the Fe K$\alpha$ line.

\citet{Bassani99} proposed  a two-dimensional diagnostic  tool to test
if an AGN is a reflection-dominated Compton-thick object, using the EW
of FeK$\alpha$ and the extinction-corrected flux ratio ($T$ parameter)
of   $F_{X}$  to   [OIII]$\lambda$5007.   Our   sample  is   shown  in
Figure~\ref{T_EW_FeKa}  where  all  available X-ray  observations  are
included.  Objects with  available $T$ and  EW(FeK$\alpha$) data 
show $T$  $\ge$ 1  and EW(FeK$\alpha$)  $\lesssim$ 1  keV, i.e.,
outside  the  locus  of  the  Compton-thick  region.   The  result  is
summarized in Table~\ref{mis_classification}.

\begin{figure}
\epsscale{.90}
\plotone{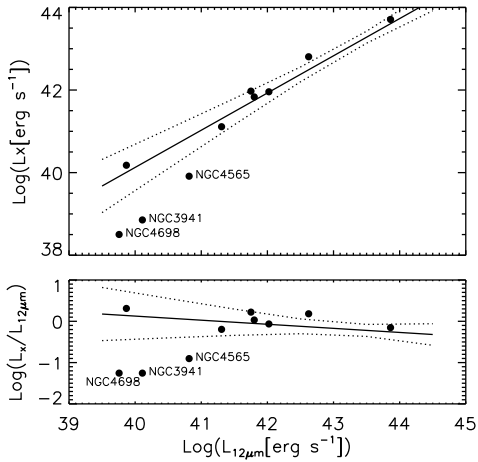}
\caption{\label{LMIR_Lx} Distribution of X-ray unobscured type 2 AGN candidates  in the plane of 
$L_{\rm MIR}$-$L_{\rm X}$. The solid line and dashed lines are the correlation and 
associated 3-$\sigma$ scatter found by \citet{Gandhi09}. }
\end{figure}

A dusty  torus in an AGN reprocesses  absorbed optical/UV/X-ray energy
and re-radiates it in the IR.  Under this process, the intrinsic X-ray
emission is expected to correlate with the IR emission.  Over the past
decade,  a  strong  IR/X-ray  correlation has  emerged  \citep{Lutz04,
  Horst08,  Gandhi09}.   We  will   use  it  to  search  for  possible
Compton-thick objects within our sample of X-ray-unobscured type 2 AGN
candidates\footnote{Although the X-ray to  the total IR flux ratio has
  been used  to identify the  Compton-thick objects in \citet{Panessa02},
  the significant contribution from stars and star formation to the IR
  emission    complicates    the    use   of    their    diagnostic.}.
Figure~\ref{LMIR_Lx}  shows the  locus of  our sample  in  the $L_{\rm
  X}$-$L_{12{\mu}m}$  plane, where the  X-ray luminosity  is corrected
for  obscuration  assuming  it  is  Compton-thin and  the  nuclear  IR
luminosity is  obtained by subtracting the  stellar and star-formation
contributions from  the total IR  emission.  The solid line  shows the
correlation  for local  Seyfert  galaxies with  well-resolved IR  cores
\citep{Gandhi09}.    As   shown   in   the   figure,   most   of   the
X-ray-unobscured type  2 candidates have the levels  of X-ray emission
expected  from their  nuclear IR  emission, confirming  that  they are
Compton-thin.  There are three exceptions,  NGC 3941, NGC 4565 and NGC
4698, with X-ray luminosities 10  times lower than those expected from
the $L_{\rm  X}$-$L_{12{\mu}m}$ correlation, given  their IR fluxes.
Being Compton-thick provides a natural explanation for their low X-ray
emission.   However,  this   explanation  lacks  evidence  from  other
aspects. First, the X-ray spectra  do not show Fe K$\alpha$ lines with
high EWs. The $F_{\rm X}/F_{\rm [OIII]}$ ratios are about 1, 4 and 0.5
for  NGC 3941,  NGC  4565 and  NGC  4698, respectively,  which is  not
significantly below  the dividing line  ($F_{\rm X}/F_{\rm [OIII]}$=1)
between Compton-thick and  Compton-thin objects.  Furthermore, the IRS
spectra  of  these  three  objects  do not  show  silicate  absorption
features,  unlike  the  general behavior  of  X-ray-obscured  objects
\citep{Shi06}. These sources  may have either intrinsically weak  X-ray emission or
very  dusty tori  that boost  their IR  emission. Due  to the  lack of
definitive explanations for their  low X-ray emission, we exclude them
from the following discussion.

After  excluding  Compton-thick  objects  as  indicated  by  extreme  low
X-ray-to-IR ratio  and intermediate type (type 1-1.9)  objects (with BELs),  
the  remaining sample  of
potential X-ray-unobscured  type  2  AGNs  is  composed of  five  objects:  IRAS
01428-0404,   NGC   3147,  NGC   4594,   NGC   7590  and   $[$H2000$]$
213115.90-424318.9, as listed  in Table~\ref{SAMPLE_TRUE}.  Due to the
observational complexity  in finding  real  unobscured  type 2
AGNs, we discuss various  possible limitations to  this final
sample further.

\subsection{The Final Sample of Possible X-ray-Unobscured Type 2 AGNs}\label{GENUINE_AGNs}
\subsubsection{The Quality of The Optical Spectra}

\begin{figure}
\epsscale{.90}
\plotone{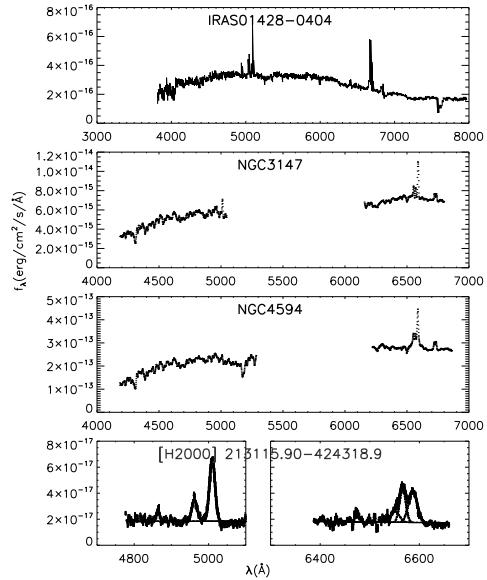}
\caption{\label{opt_spec_type2agn}  The published optical spectra for
four out of the five possible X-ray-unobscured type 2 AGNs.}
\end{figure}

Figure~\ref{opt_spec_type2agn}  shows the available  published optical
spectra for four objects. No  broad line component is visible in these
spectra. Since a so-called ``non-detection'' actually depends
on  the S/N  that the  observation achieves,  we  develop quantitative
methods to  measure the  reliability of ``no  BELs'' in  these objects
based on  the upper-limits to the  BEL fluxes. The  upper-limits on the
broad H$\alpha$ fluxes for NGC 3147 and [H2000] 213115.90-424318.9 are
from  \citet{Bianchi08a}  and  \citet{Panessa09},  respectively.   For
IRAS01428-0404  and NGC 4594  with published  flux-calibrated spectra,
the      3-$\sigma$  upper limits      are     derived      through
3$\sigma_{F_{\lambda}}$(FWHM*$\Delta\lambda)$$^{0.5}$,            where
$\sigma_{F_{\lambda}}$  is  the  noise  of  the flux  density  at  the
spectral resolution of $\Delta\lambda$. The FWHM is assumed to be 2200
km/s, the median value for the sample of  \citet{Ho97b}.\footnote{There is 
no published spectrum of NGC 7590, although it is described as being a
potentially unobscured Type 2 AGN \citep{Panessa02}.}

\begin{figure}
\epsscale{.90}
\plotone{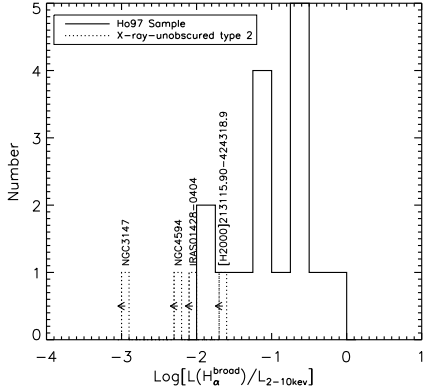}
\caption{\label{Ha_Lx_truetype2}   The  solid   histogram   shows  the
distribution of the  extinction-corrected broad H$\alpha$ luminosity
to  hard  X-ray luminosity  ratio  for  the  type 1-1.9  objects  in
the sample of \citet{Ho97b}.  The   dotted   histogram  shows   the
distribution for X-ray-unobscured type 2 AGN candidates.}
\end{figure}

\begin{figure}
\epsscale{.90}
\plotone{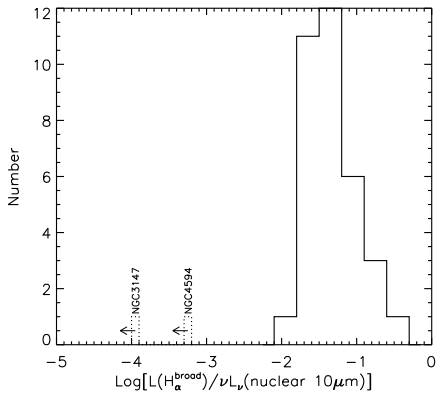}
\caption{\label{Ha_Nband_truetype2} 
The solid  histogram shows the  distribution of the  broad H$\alpha$
luminosity to 10 $\mu$m nuclear IR luminosity ratio for type 1.0-1.2
AGNs,  while  the  dotted   histogram  shows  the  distribution  for
the X-ray-unobscured  type   2  AGNs with suitable nuclear IR measurements.  The   H$\alpha$  luminosity
upper-limits at  given nuclear  IR luminosities of  these unobscured
type 2  AGNs are  significantly lower than  those of  general type-1
AGNs. }
\end{figure}

\begin{figure}
\epsscale{.90}
\plotone{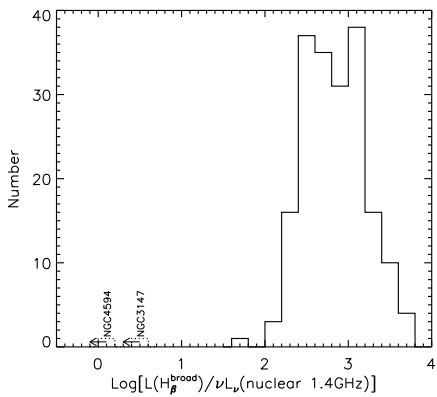}
\caption{\label{Hb_radio_truetype2} 
The solid  histogram shows the  distribution of the  broad H$\beta$
luminosity to 1.4 GHz nuclear radio luminosity ratio  from
\citet{Li08},  while  the  dotted   histogram  shows  the  distribution  for
X-ray-unobscured  type   2  AGNs (with suitable radio measurements).  The   H$\beta$  luminosity
upper-limits for the unobscured
type 2  AGNs are  significantly lower than  those of  general type-1
AGNs. }
\end{figure}

We evaluate the strength of the BELs in these galaxies (represented by
the  broad H$\alpha$ flux)  relative to  the X-ray,  mid-infrared, and
radio nuclear continuum  fluxes.  These metrics are put  in context by
comparing   them  with  the   distributions  for   unobscured,  type-1
AGN.   Figure~\ref{Ha_Lx_truetype2}    shows   the   distribution   of
extinction-corrected  broad H$\alpha$ to  hard X-ray  luminosity ratio
($L_{\rm H\alpha}^{\rm Broad}/L_{\rm X}$) for type 1-1.9 galaxies with
solid broad H$\alpha$ measurements  from \citet{Ho97b}. The hard X-ray
data  are   from  \citet{Ho01},  \citet{Terashima02},  \citet{Ptak04},
\citet{Pellegrini05},  \citet{Gonzalez-Martin06}, \citet{Cappi06}, and
\citet{Panessa06}.  A second  test is  based on  the  observation that
measurements at  wavelengths near  12$\mu$m provide a  reasonably good
estimate of  the total luminosity  of an AGN  \citep{Spinoglio89}.  We
use  small-beam measurements  of the  N-band flux  for type-1  AGN and
compare    them   with    the    broad   H$\alpha$    as   shown    in
Figure~\ref{Ha_Nband_truetype2}.    The  H$\alpha$   fluxes   for  the
comparison  sample  (all type  1.0  -  1.2)  are from  \citet{Lacy82},
\citet{Rafanelli85},     \citet{Neugebauer79},    \citet{Ho97b}    and
\citet{Morris88}.   The  $L_{\rm  Nband}$  are the  N-band  (10$\mu$m)
luminosities   measured  through  small   apertures  (a   diameter  of
$\sim$2$''$   -  6$''$)  and   are  collected   from  \citet{Rieke78},
\citet{Neugebauer79},   \citet{Heckman83},   \citet{Maiolino95b}   and
\citet{Gorjian04} \footnote{Small-beam N-band and 1.4 GHz measurements
  are      not     available      for     IRAS      01428-0404     and
  [H2000]213115.90-424318.9.}.   The  third   test  is  based  on  the
correlation between  radio luminosity and the  broad H$\beta$ emission
as seen by \citet{Li08}, as shown in Figure~\ref{Hb_radio_truetype2}, where
the H$\beta$ flux is taken to be H$\alpha$/3.1.

We now  discuss  five objects that possibly have
relatively extreme behavior according to the above three metrics.

{\bf NGC  3147}: NGC  3147 has upper limits  on the  ratio of  the broad
H$\alpha$ and H$\beta$ lines to  X-ray, nuclear IR  and radio luminosities 2-3  orders of
magnitude  lower than the  average value  of general  type 1  AGNs, as
shown in Figures~\ref{Ha_Lx_truetype2}$\textendash$\ref{Hb_radio_truetype2}.
All three  tests indicate that NGC  3147 is the  strongest case where
the BEL  is absent or  extremely weak.  The upper-limit  on the
BEL  flux  is based  on  an  assumption of  a  2000  km s$^{-1}$  FWHM
\citep{Bianchi08a}  and a strength larger by 2-3 orders  of magnitude  would be
required  to  bring  this  galaxy   up  into  the  normal  range. Only if the
line width were far greater than we have assumed would a significant amount
of BEL flux be missed. However, given the relation between virial
mass around a black hole and the H$\alpha$ and H$\beta$ line widths \citep{Denny09}, 
plus the moderate stellar mass in NGC 3147 (M$_K \sim -23.8$) and thus 
the expected moderate black hole mass (Table~\ref{SAMPLE_TRUE}), such an explanation would be contrived.

{\bf NGC  4594}:  NGC 4594  is  the  second strongest  case  for  no BEL  as
indicated by Figures~\ref{Ha_Lx_truetype2}$\textendash$\ref{Hb_radio_truetype2}.
\citet{Kormendy96} suggested the presence  of broad H$\alpha$ based on
the  overall  profile   in  a  high-spatial-resolution  HST  spectrum.
However, \citet{Nicholson98} argued that  the line profile arises from
a  combination of  three  narrow lines,  H$\alpha$ and  [NeII]$\lambda
\lambda$6548, 6583.  Walsh  et al.  (2008) instead found  from more recent HST
observations that the fit  was improved including broad H$\alpha$, but
were not  able to  reach a definite  conclusion because the  signal to
noise  was not  high  enough. On  the  other hand,  no broad  H$\beta$
component is  seen \citep{Kormendy96}. A broad component  1-2 orders of
magnitude  brighter  than  the  upper  limit we  have  used  would  be
necessary to  bring this  galaxy into the  range of normal  broad line
strengths (Figures 7 and 8). The stellar mass in this galaxy again
implies only a moderately massive black hole (Table~\ref{SAMPLE_TRUE}), 
so as for NGC 3147, it is unlikely that the lines are so broad that they have beenmissed.

{\bf IRAS01428-0404} and {\bf [H2000]  213115.90-424318.9}: The upperlimits on the
$L({\rm H}_{\alpha}^{\rm broad})$/$L_{\rm 2-10kev}$  are at the edge of
the  distribution  of  general  type  1  AGNs.   Deeper  spectroscopic
observations  with small slits  are required to test whether 
these galaxies harbor extremely faint BELs.
 
{\bf NGC  7590}: As there is  no  available published  optical spectrum, we cannot comment on how reliable the absence of a BEL is \citep{Vaceli97}.

In conclusion, of these five remaining possible X-ray-unobscured type 2 AGNs,  two 
(NGC 3147 and  NGC 4594) have optical spectra  with sufficient S/N to
indicate extremely weak or absent BELs. For  the remaining three,
there is only moderate evidence that they may harbor no or extremely
weak BELs.

\subsubsection{The Aperture for the X-ray Observations}

Our identification  of unobscured X-ray emission in  these objects may
be mistaken  if the  X-ray emission is  actually dominated  by diffuse
and/or  extra-nuclear point  source emission  that shows  little X-ray
obscuration.   Some  examples are  shown  in \citet{Brightman08}.   An
example in this study is NGC  4698. It shows a large difference in the
X-ray flux between the small  aperture of {\it Chandra} and large ones
of other observatories.  It is  classified as an unobscured type 2 AGN
based  on  the  ASCA  observations  \citep{Panessa02}.   As  shown  in
\S~\ref{COMPTON_AGNs} and Figure~\ref{LMIR_Lx},  this object is likely
Compton-thick based on the flux  observed with {\it Chandra} and it is
excluded from our  final sample.  Only one (NGC 3147) of  the final sample listed
in Table~\ref{SAMPLE_TRUE}  has a high  spatial-resolution (resolution
of 0.5$''$) {\it  Chandra} observation.  Extra-nuclear contamination 
could in principle be an explanation for the unusual properties of the remaining 
objects, but it is  not  very likely.   As
indicated  in Table~\ref{sample_X-ray},  only one  object (not  in our
final sample) suffers from this problem among $\sim$10 objects with
multiple X-ray observations in different apertures.

\subsubsection{AGN Variability}

AGNs can change  their optical classification on a  timescale of years
to decades. Variation in X-ray  obscuration has also been observed for
many objects  \citep[e.g.][]{Matt03, Elvis04}, either due  to a change
in the  properties of  the nucleus  or in the  motion of  an obscuring
cloud intercepting our line of  sight.  A fake X-ray-unobscured type 2
AGN could  result from optical  and X-ray variability, if  observed at
different   epochs.    However,   simultaneous   optical   and   X-ray
observations have been  carried out for NGC 3147  and exclude variable
obscuration  as  a possibility  for  its behavior  \citep{Bianchi08a}.
Similar constraints  are available for  $[$H2000$]$ 213115.90-424318.9
\citep{Panessa09}. For NGC 4594, multiple X-ray observations do not detect
variability and multiple sets of optical data do not reveal BEL variations,
but the measurements are not truly simultaneous. Objects  still  lacking  simultaneous  X-ray  and
optical    observations    are   indicated    in    column   (4)    of
Table~\ref{SAMPLE_TRUE}.

\section{Discussion}\label{discussion}

\begin{figure}
\epsscale{.90}
\plotone{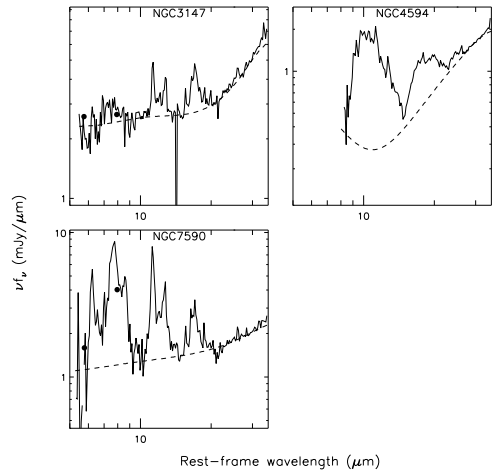}
\caption{\label{SPEC_GENUINE} Stellar-photosphere-subtracted IRS spectra (solid curves) of
X-ray-unobscured type 2 AGNs, where the filled
circles are IRAC data. The dashed line is the fitted continuum. }
\end{figure}

IR spectra of the two weak-BEL AGN, NGC 3147 and 4594, plus the unconfirmed candidate 
NGC 7590 are  shown in Fig.~\ref{SPEC_GENUINE}, while the remaining
two (IRAS01428-0404 and [H2000]213115.90-424318.9) were not observed by
{\it Spitzer}.  NGC 7590  has strong aromatic emission indicating its
mid-IR output is dominated by star forming regions, while NGC 3147 and 4594
show strong contributions by the AGN. We will now put the infrared
properties of these AGNs into context with other observations.

\subsection{NGC 3147: The Prototypical Unobscured Type 2 AGN}

NGC  3147 presents  the most  complete case  for having  an unobscured
Type-2  AGN, as listed  in Table~\ref{SAMPLE_TRUE}.   Since it  can be
considered a prototype for these  objects, we discuss its status first.

NGC 3147 was found to have  an unobscured X-ray spectrum and no BEL in
simultaneous observations  by \citet{Bianchi08a}.  \citet{Brightman08}
carried out a  more detailed analysis of the  X-ray data and concluded
that  a  Compton-thick component  could  not  be completely  excluded,
although  they  pointed  out   that  the  X-ray  variability  provided
substantial  evidence against  this  possibility. This  can be  tested
further  with our  infrared  spectrum.  In addition to aromatic
features of modest equivalent width, a substantial power-law continuum is present. The relatively
strong feature in the 16 - 19 $\mu$m region is probably due in part to silicates
in emission. The spectrum  shows no  silicate absorption  as would  be expected  from a
heavily obscured  AGN \citep[e.g.][]{Shi06}.  Together  with the X-ray
variability, the IR spectrum makes the Compton-thick possibility very
unlikely.

\begin{figure}
\epsscale{.90}
\plotone{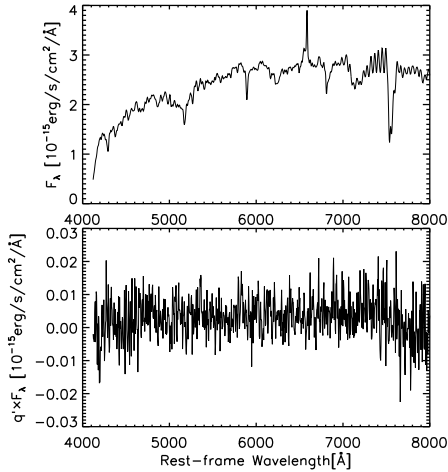}
\caption{\label{NGC3147_SPEC} The total flux density (upper panel) and the polarized flux 
density (lower panel) for NGC 3147. The
linear Stokes parameter q$'$ has been rotated so that the polarization position angle is 
zero averaged over the entire spectrum.}
\end{figure}

\citet{tran05} report  that spectropolarimetry of this  galaxy did not
reveal a hidden  BEL, but no details of  the observation are provided.
The  optical spectropolarimeteric  observation  for NGC  3147 in  this
study  is shown  in  Figure~\ref{NGC3147_SPEC}, and  H$\alpha$ is  not
detected in  polarized light. We  set a 3-$\sigma$ upper-limit  to the
peak  of   any  polarized  H$\alpha$   line  of  1.4$\times$10$^{-17}$
erg/s/cm$^{2}$/$\AA$.   The  two  results  together provide  a  strong
limit,  since they  show  that  there were  no  polarized broad  lines
detected at  two separate epochs.  
 
Despite the evidence in the X-ray  and mid-infrared for a low level of
obscuration,    as   shown    in    Figure~\ref{Ha_Lx_truetype2}$\textendash$\ref{Hb_radio_truetype2}, NGC  3147 has  an upper limit  to its
broad H$\alpha$ more than two  orders of magnitude lower than would be
expected for  a type-1  AGN of its  intrinsic nuclear  luminosity. The
very stringent  limit on any scattered broad  H$\alpha$ emission makes
it difficult to appeal to some unique geometry to hide the BEL region.
Our metrics could be deceptive if the nuclear emission were dominated by
star formation, since that would strengthen all the continuum luminosity
indicators relative to the broad H$\alpha$. However, the mid-infrared
spectrum of NGC 3147 shows only a modest level of contribution from
the aromatic bands, indicating that the AGN is dominant. 
Thus, the case  is very strong that this  galaxy lacks any significant
level of broad line emission.

The case for NGC 4594 having anomalously weak broad lines is also
strong. Like NGC 3147, it falls far below the normal Seyferts in its
broad H$\alpha$ strength relative to the AGN luminosity indicators, and
its mid-IR spectrum is dominated by the AGN. Although it does not have
X-ray measurements simultaneously with spectroscopy, repeated measurements
of both types do not indicate variability at a level that could account
for its apparently weak H$\alpha$. As
shown in  column (13)  of Table~\ref{SAMPLE_TRUE}, a variable UV core is detected
in NGC 4594. Due to its high sensitivity  to 
dust obscuration, the detection of the
nuclear  UV  emission  and  its  variability  imply a low level of obscuration for this AGN.

\subsection{Candidate Unobscured Type-2 AGNs as a Whole}

With NGC 3147 and NGC 4594 as templates, we review the behavior of the
candidates most likely to represent additional weak-BEL AGNs. 

\subsubsection{Silicate Features}

We previously excluded NGC 3941,  NGC  4565 and  NGC  4698 from consideration
because their X-ray properties were between those of Compton thick and
unobscured AGN. As shown in Figure~\ref{OBS_SPEC}, the infrared spectra
of all three galaxies show silicate emission and resemble the spectrum of
NGC 4594 closely. This behavior supports the possibility that these objects 
may be unobscured with somewhat peculiar X-ray properties. Additional
measurements in the X-ray may elucidate whether they are weak-BEL AGN. 
In comparison, the strong aromatic features in the spectrum of NGC 7590
indicate powerful nuclear star formation that might cause our metrics to
underestimate the intrinsic BEL strength relative to the continuum emission
of the AGN. 

For  typical  AGNs,   the  existence  of  the  dusty   tori  has  been
demonstrated  extensively through  thermal  IR continuum  observations
\citep[e.g.][]{Neugebauer79}, the discovery  of polarized BELs in type
2 AGNs \citep{Miller83,  Hines95, Moran00, Hines99, Tran01, Schmidt02}, detection
of  the  silicate   features  \citep{Roche91,  Siebenmorgen05,  Hao05,
  Weedman05,  Shi06}, and  interferometric imaging  of  nearby Seyfert
galaxies \citep{Jaffe04, Tristram07}.  In the unification model, these
tori are responsible  for obscuring the type-1 nuclei  to yield type 2
AGNs. The detection  of the silicate feature in emission is evidence
for the existence of dusty tori or similar structures in the weak-BEL
AGNs. It is possible that the dusty tori in these peculiar sources 
have special properties to account for their unusual nature.

The  material in  the dusty  torus is  either smoothly  distributed or
clumpy     (cloudy)      \citep{Fritz06,     Elitzur08,     Nenkova08,
  Hatziminaoglou09}.  In  the  smooth  model, the  UV/X-ray  radiation
directly heats the inner dusty  wall that shields the outer part.  The
dust temperature generally  decreases monotonically with distance from
the  central  accretion  disk.  Therefore,  while the  BEL  region  is
obscured by the dusty torus, the  inner hot wall is also invisible and
obscured by the  outer cold part. In this  case, a silicate absorption
feature should be present, i.e., the smooth torus model cannot explain
the simultaneous presence of the  silicate emission feature and a type
2 optical spectrum in these AGNs.

The smooth dusty  model may over-simplify the real  situation  for
AGNs  in general.   For example,  the Seyfert-2/Seyfert-1  ratio  of 4
indicates a large covering factor and thick torus \citep{Maiolino95a},
as is also found by  interferometric imaging  of nearby  Seyfert galaxies
\citep{Jaffe04, Tristram07}.  A thick smooth model is also dynamically
unstable  against gravitational  force.  Consequently,  several models
have   been    proposed   based   on   a    thick   cloudy   structure
\citep[e.g.][]{Krolik07, Schartmann09}.  Clumpy models can explain the
Compton-thick/Compton-thin transitions  in some objects \citep{Matt03,
  Elvis04} in terms  of a Compton-thick cloud moving  through the line
of  sight.  In general for AGNs, the  transition  of the  silicate feature  from
emission to absorption with  increasing X-ray obscuration supports the
predictions of the clumpy  model \citep{Shi06, Schartmann09}.  For the
unobscured type 2 AGNs,  the silicate emission is therefore consistent
with  the  small level  of  X-ray  absorption  \citep[see Fig.   3  of
][]{Shi06}.   Both characteristics  indicate that  the dusty  torus is
viewed face-on, i.e., it is not  likely that a dusty torus can obscure
any BEL region if it exists in these AGNs.

The  co-existence of  the silicate  emission feature, the type  2 optical
spectrum and the small X-ray obscuration is therefore not  compatible with standard
proposals for the circumnuclear tori in these AGNs.  It has been suggested
that   their tori   may   have   a  higher-than-normal   dust-to-gas   ratio
(i.e. $A_{\rm  V}/N_{\rm H}$), leading  to much higher  obscuration of
the BEL  region than  that of the  X-ray emission. However,  since the
torus must  still be  viewed edge-on to  obscure the BEL  region, this
model  is inconsistent  with  the presence  of  the silicate  emission
feature.

\subsection{Theoretical Interpretation For Weak BELs}

\begin{figure}
\epsscale{.90}
\plotone{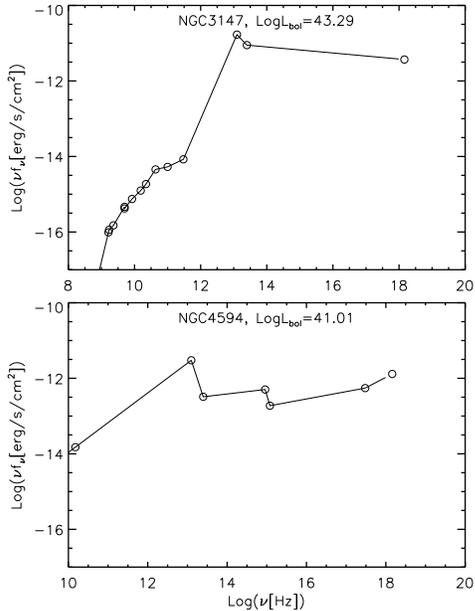}
\caption{\label{BOLOMETRIC_SED} The broad-band  nuclear SED from X-ray
  to radio  of X-ray-unobscured  type 2 AGNs.  The IR bump between $10^{12}$
and $3 \times 10^{13}$ Hz is present
  while the UV bump near $10^{15}$ Hz is weak or absent (compare also Figure 11 for NGC 3147).}
\end{figure}

There are several theoretical  works that predict the disappearance of
the  BEL region  under  certain bolometric  luminosities or  accretion
rates \citep{Elitzur06,  Czerny04, Laor03, Nicastro00,  Elitzur09}. In
order to  compare our  sample to the  predictions of these  models, we
have  compiled the  black-hole  (BH) masses  from  the literature  and
calculated the  bolometric luminosities. For NGC 3147 and
NGC   4594,  we   have  compiled   the  nuclear   SED  as   shown  in
Figure~\ref{BOLOMETRIC_SED}                                         and
Tables~\ref{NSED_3147}$\textendash$\ref{NSED_4594}.   The  nuclear  IR
photometry at 12 $\mu$m and  24 $\mu$m was obtained by subtracting the
stellar and star formation contribution  from the total IR emission as
described  in  \S~\ref{IR-DATA}.   Fluxes  at other  wavelengths  were
obtained  through  high   spatial-resolution  observations.   We  have
integrated  the  SED from  $10^{8}$  GHz  to  10 KeV;  the  bolometric
luminosities are listed in Table~\ref{SAMPLE_TRUE}.  For the remaining
objects, the X-ray flux is used to calculate the bolometric luminosity
using  the  luminosity-dependent   relation  of  X-ray  to  bolometric
luminosity \citep{Shankar04}.

\begin{figure}
\epsscale{.90}
\plotone{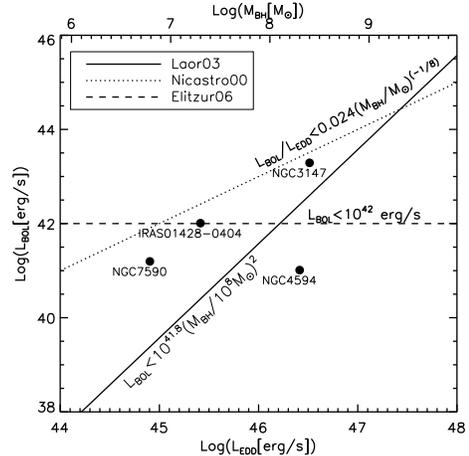}
\caption{\label{LBol_LEDD} Distribution of  X-ray-unobscured type 2 AGNs from Table 6 in the
plane of bolometric luminosity and Eddington luminosity (BH mass). The three lines are the three
theoretical upper-limits for the absence of the BEL region. }
\end{figure}

In  the disk-outflow  model \citep{Elitzur06},  the clouds  across the
accretion disk rise  into the wind and move  outward, forming both the
BEL  and  dusty torus  regions.   The  inner  hot ionized  clouds  are
responsible for the observed BELs while the outer dusty clouds are the
components of the clumpy torus.  The outflow rate of clouds is related
to  the  available material  reservoir  in  the  accretion disk.   For
$L_{\rm BOL}$  $<$ 10$^{42}$ erg  s$^{-1}$, the accretion rate  is not
high  enough  to provide  enough  cold  clouds  and the  clumpy  torus
disappears.  At somewhat lower  accretion rates, the hot cloud outflow
is further  suppressed, resulting in  a vanishing of the  BEL regions.
As  shown  in  Figure~\ref{LBol_LEDD},   NGC  4594  has  a  bolometric
luminosity below  10$^{42}$ erg s$^{-1}$.  However,  this galaxy shows
strong  silicate  emission  features   at  both  9.7  and  18  $\mu$m,
indisputable    evidence   for    the   existence    of    the   dusty
torus. \citet{Elitzur09}  have derived a  lower bound below  which the
BEL does  not exist, which is  even more difficult  to make compatible
with our results.

There is a limit on the maximum velocity of the BEL ($\sim$ 25, 000 km
s$^{-1}$)  \citep{Laor03}.  Using  the well-known  correlation between
the BEL  size and the  luminosity, such a  limit implies BELs  may not
exist  at  low  luminosities  for  a given  BH  mass.   This  critical
luminosity  is  approximately 10$^{41.8}$($\frac{M_{BH}}{10^{8}})^{2}$
erg   s$^{-1}$.    Only   one   object   is  below   this   limit   in
Figure~\ref{LBol_LEDD}.     Although   the    maximum    velocity   is
observationally uncertain, to include most of our sample in the region
where  \citet{Laor03}  would  predict  no BEL,  the  required  maximum
velocity needs to be ten  times lower, obviously inconsistent with the
observations.

\citet{Nicastro00} proposes  a vertical outflow  as the origin  of the
BEL clouds, in  a configuration where the radially  accreting disk and
vertically  outflowing corona  coexist.   Below an  accretion rate  of
$\dot{M}/\dot{M}_{\rm
  EDD}$=$0.3\eta({\alpha}\frac{M}{M_{\odot}})^{-1/8}$,  the  accretion
disk  is  gas-pressure-dominated throughout.   As  a  result, all  the
available energy is dissipated  in the disk and no radiation-supported
and -driven  wind is  produced.  Adopting a  mass-to-energy conversion
coefficient $\eta$=0.06 and  a viscosity coefficient $\alpha$=0.1, the
limiting  bolometric luminosity  is  shown in  Figure~\ref{LBol_LEDD}.
This  model  can explain  all  of the  objects  in  our sample.   This
suggests that the luminosity and  accretion rate may play an important
role in the existence of the BEL region. However, the situation may be
more complicated as there are  some type 1 objects with low luminosity
($<$10$^{42}$  erg s$^{-1}$)  and accretion  rate ($<$10$^{-3}$$L_{\rm
  EDD}$) \citep{Ho08, Elitzur09}.

\section{CONCLUSION}\label{conclusion}

We have presented a multi-wavelength study of unobscured type
2 AGNs. Our conclusions are the following:

(1) We  have found that the  original sample of this  proposed type of
AGN  is contaminated by  many objects  with BELs,  as revealed  by our
consistent optical  classification and new observations.  One of these
objects, RXJ  1737.0+6601, has  a highly polarized  optical continuum.
Additional contaminants  include several new  Compton-thick candidates
with extremely low nuclear X-ray-to-IR ratios.

(2) We have identified two objects that appear to be true unobscured type-2
AGNs: NGC 3147 and NGC 4594. They have little  X-ray extinction  with $N_{\rm
  H}$  $<$  $\sim$10$^{21}$ cm$^{-2}$.  The  upper-limits  on the  BEL
luminosities at a  given nuclear X-ray, IR or  radio luminosity are two
orders of magnitude lower in relative BEL strength than the average of  typical type-1 AGNs. 
Several other galaxies remain as candidates to be weak-BEL AGN. 

(3) From the small number of confirmed cases, unobscured Type 2 AGN do
exist but they are very rare.

(4) The IR spectra of the unobscured  type 2 AGNs and of many of the candidates show silicate
emission features. The presence of the silicate features
 demonstrates  the existence of dusty tori and that the tori are viewed approximately face-on.

(5) Thus, in contradiction to  the simple unified model, the X-ray and
IR  properties indicate that  the nuclei  are viewed  directly without
intervening obscuring  material, despite the  intrinsically weak broad
emission lines.

(6) The  distributions of the bolometric luminosity  and the accretion
rate  of these  objects contradict  some theoretical  studies  but are
consistent with the work of \citet{Nicastro00}.

\acknowledgements

We thank the  anonymous referee for detailed comments.  Support
for this work was provided  by NASA through contract 1255094 issued by
JPL/California Institute of Technology.  This research has made use of
the NASA/IPAC  Extragalactic Database (NED)  which is operated  by the
Jet Propulsion  Laboratory, California Institute  of Technology, under
contract with the National Aeronautics and Space Administration.  This
research has made use of the NASA/IPAC Infrared Science Archive, which
is operated by the  Jet Propulsion Laboratory, California Institute of
Technology,  under contract  with the  National Aeronautics  and Space
Administration.

\clearpage


\tabletypesize{\scriptsize}
\begin{deluxetable}{lllcccclllll}
\tablecolumns{15}
\tablecaption{\label{sample_opt} Optical Properties of X-ray-Unobscured Type 2 AGN Candidates}
\tablehead{
\colhead{sources}                             & 
\colhead{D}                                   & 
\colhead{Type}                                & 
\colhead{$F_{\rm [O III]{\lambda}5007}$}                              &
\colhead{$\frac{{\rm [NII]{\lambda}6583}}{{\rm H}\alpha}$}         &
\colhead{$\frac{ {\rm O[III]}{\lambda}5007} {{\rm H}{\beta} }$ }   &
\colhead{$\frac{{\rm [O I]\lambda6300}}{{\rm H}\alpha}$}           &
\colhead{$\frac{{\rm [S II]\lambda6716, 6731}}{{\rm H}\alpha}$}    &
\colhead{$F_{\rm Broad H_{\alpha}}$ }                                 & 
\colhead{Ref.} \\
\colhead{(1)}                                 &
\colhead{(2)}                                 &
\colhead{(3)}                                 &
\colhead{(4)}                                 &
\colhead{(5)}                                 &
\colhead{(6)}                                 &
\colhead{(7)}                                 &
\colhead{(8)}                                 & 
\colhead{(9)}                                 &
\colhead{(10)}                               
}
\startdata
IRAS 00317-2142  &    113.2         & HII+BEL(H$\alpha$)          &  10.2  & 0.45      &     0.93  &  0.03   &  0.21       &  8.51    &    1 \\
IRAS 01428-0404  &    75.6          & S2                          &  --    & 0.7       &     3.7   &  0.2    &  0.4        &   --     &    2 \\
Mrk 993          &    65.1          & S1.5                        &  15.4  & 2.20      &    12.02  &  0.46   &  1.35       &   --     &    3 \\
NGC 3147         &    45.1          & S2                          &  9.50  & 2.71      &     6.14  &  0.15   &  1.14       &   --     &    4 \\
NGC 3660         &    58.5          & HII+BEL(H$\alpha$)          & 98.8   & 0.47      &     2.80  &  0.05   &  0.20       &  2.0     &    5,6 \\  
NGC 3941         &    12.2$^{a}$    & S2                          &  3.60  & 1.56      &     3.52  &  0.130  &  0.95       &          &    4  \\
NGC 4565         &    17.5$^{a}$    & S1.9                        &  6.87  & 2.50      &     8.73  &  0.31   & 0.97        &   --     &    4  \\
NGC 4579         &    14.5          & S1.9/L1.9                   &  8.88  &  1.89     &     3.07  &  0.48   & 1.52        &   --     &    4  \\
NGC 4594         &     9.8$^{a}$    & L2                          &  6.65  &  2.19     &     1.57  &  0.18   & 1.07        &   --     &    4  \\
NGC 4698         &    14.3          & S2                          &  2.15  &  1.31     &     4.29  & 0.074   & 0.88        &   --     &    4  \\
NGC 5033         &    18.0          & S1.5                        & 18.1   &  2.36     &     4.69  & 0.29    & 1.08        &   --     &    4  \\
H1320+551        &    284.6         & S1.8                        & 110    &  1.40     &     18.17 &  0.36   & 0.72        &  --      &    7  \\
Mrk 273x         &   1964.8         & S2                          &  0.138$^{b}$ &     & 5.75$^{b}$ &         &            &   --     &     8  \\
NGC 5995         &    117.5         & S1.9/HII+BEL(H$\alpha$)     & 660    &  1.02     &     6.31  & 0.058   & 0.22        &   --     &    9  \\
NGC 6251         &    109.1         & S1                          &  4.4   &  2.65     &    11.21  & 0.25    & 0.29        &   --     &    10 \\
IRAS 20051-1117  &    138.7         & HII+BEL(H$\alpha$, H$\beta$)&  6.11  &  0.61     &     2.58  & 0.054   & 0.32        & 2.63     &    1  \\
NGC 7590         &    22.9          & S2                          &  3.0   &  1.05     &     5.0   & 0.11    & 0.86        &   --     &    11 \\
NGC 7679         &    72.8          & HII+BEL(H$\alpha$)          &  --    &  0.50     &     0.95  & 0.035   & 0.28        & 5.86     &    12, 13 \\
RXJ1715.4+6239   &   3642.1         & QSO2.0?                     &  --    &  --       &     --    &  --     &  --         &   --     &    14 \\
RXJ1724.9+6636   &   2910.3         & QSO2.0?                     &  --    &  --       &     --    &  --     &  --         &   --     &    14 \\
RXJ1737.0+6601   &   1534.8         & QSO1.0                      &  --    &  --       &     --    &  --     &  --         &   --     &    15  \\
XBSJ031146.1-550702 & 775           & S1.9                        &  --    &  --       &     --    &  --     &  --         &   --     &    16 \\ 
$[$H2004$]$213004.24-430744.2 & 1349    & S1.2                        & 0.18   &  1.69     &    10.64  &  --     &  0.51       &   --     &    17 \\
$[$H2000$]$213115.90-424318.9 & 1947    & S2                          & 0.31   &  1.33     &    8.97   &  --     &  --         &   --     &    17 \\
\enddata
\tablecomments{ Col.(1):  Sources; Col.(2):  Distance in Mpc  given by
D(Virgo+GA+Shapley)  as  measured by  NED  for H$_{0}$=70  km/s/Mpc,
$\Omega_{\rm  m}$=0.3 and $\Omega_{\lambda}$=0.7. $^{a}$  Distance is from \citet{Tonry01}.  Col.(3): Optical
AGN  type.  Col.(4):  Extinction-corrected [OIII]$\lambda$5007  line
flux in the unit of 10$^{-14}$ erg/cm$^{2}$/s.   The  extinction  is  derived  from  Balmer  decrement  using
$E(B-V)$   =    2.02${\rm   log}(R/R_{\rm   intr})$    and   $A_{\rm
  [OIII]5007\AA}=3.47E(B-V)$  based  on   the  extinction  curve  of
\citet{Fitzpatrick99}    (see   their    Fig.6),    where   $R$    =
H$_{\alpha}$/H$_{\beta}$  and  $R_{\rm  intr}$ =  3.1. $^{b}$ Not corrected 
for extinction due to the lack of H$\alpha$ measurement. Col.(5)-(8):
Extinction-corrected      $\frac{{\rm      [NII]{\lambda}6583}}{{\rm
    H}\alpha}$,  $\frac{ {\rm O[III]}{\lambda}5007}  {{\rm H}{\beta}
}$, $\frac{{\rm  [O I]\lambda6300}}{{\rm H}\alpha}$  and $\frac{{\rm
    [S   II]\lambda6716,  6731}}{{\rm   H}\alpha}$.    Col.(9):  The
observed broad H$\alpha$ flux in the unit of 10$^{-14}$ erg/cm$^{2}$/s.  Col.(10): References for optical line
fluxes.\\
References: 1 -- \citet{Moran96}; 2 -- \citet{Pietsch98}; 3 -- \citet{Corral05}; 
4 -- \citet{Ho97a}; 5 -- \citet{Kollatschny83}; 6 -- \citet{Goncalves99}; 7 -- \citet{Barcons03}; 
8 -- \citet{Xia99}; 9 -- \citet{Lumsden01}; 10 -- \citet{Ferrarese99}; 11 -- \citet{Vaceli97}; 
12 -- \citet{Kewley01}; 13 -- \citet{DellaCeca01}; 14 -- \citet{Wolter05}; 15 -- This work; 
16 -- \citet{Caccianiga04}; 17 -- \citet{Panessa09} }
\end{deluxetable}
 \clearpage

\begin{deluxetable}{llllllllllllllll}
\tabletypesize{\scriptsize}
\tablewidth{0pc}
\tablecaption{\label{POL_RESULT} Polarization Observations}
\tablehead{     
\colhead{sources}     &
\colhead{Wavelength}  &
\colhead{Polarization}&
\colhead{P.A.}         }
\startdata
RXJ 1737.0+6601     &  [3500, 6000]\AA & 2.41$\pm$0.08\%  & 7.5$\pm$0.9$^{\circ}$ \\
NGC3147             &  [4500, 7500]\AA & 0.14$\pm$0.01\%  & 158.1$\pm$1.6$^{\circ}$ \\
NGC4698             &  [4500, 7500]\AA & 0.13$\pm$0.01\%  & 167.8$\pm$1.7$^{\circ}$ \\
\enddata
\tablecomments{The error bar is given at 1-$\sigma$ level.}
\end{deluxetable}

\begin{deluxetable}{llllclllllll}
\tabletypesize{\scriptsize}
\tablewidth{0pc}
\tablecaption{\label{sample_X-ray} X-ray Properties of X-ray-Unobscured Type 2 AGN Candidates}
\tablehead{     \colhead{sources}              &  \colhead{Observatory}            &   \colhead{$f_{2-10 keV}$}   & 
                \colhead{$N_{\rm H}$}           &  EW$_{\rm FeK{\alpha}}$            &    \colhead{Ref.}\\
                \colhead{}                     &  \colhead{}                       &    \colhead{[10$^{-12}$erg/s/cm$^{2}$]}     & 
                \colhead{[10$^{20}$cm$^{-2}$]}  & \colhead{[eV]} \\
\colhead{(1)}                                 & 
\colhead{(2)}                                 & 
\colhead{(3)}                                 & 
\colhead{(4)}                                 & 
\colhead{(5)}                                 & 
\colhead{(6)}                                 & 
} 
\startdata
IRAS 00317-2142  & Chandra & 0.21          &    8$\pm$3              & $<$ 1400               & 1  \\
                 & ASCA    & 0.8           &    1.9                  & $<$ 900                & 2  \\
IRAS 01428-0404  & ASCA    & 0.4           & 32$^{+109}_{-32}$        & --                     & 3  \\
Mrk 993          & XMM     & 1.34          &    7.2$\pm$1.1          & $\sim$270              & 4  \\
NGC 3147         & Chandra & 3.7           &    14.8$^{+3.7}_{-1.8}$  &    --                  & 5  \\                 
                 & XMM     & 1.48$\pm$0.07 &    2.8$\pm$1.2          &    130$\pm$80          & 6  \\
                 & SAX     & 2.2           & $<$2.9                  &    675$^{+395}_{-328}$  & 3  \\ 
NGC 3660         & ASCA    & 2.3           & $<$3                    & --                     & 7  \\
NGC 3941         & XMM     & 0.04          & $<$10                   & --                     & 8  \\
NGC 4579         & Chandra & 5.2           &  --                     &   --                   & 9  \\
                 & XMM     & 3.85          &$\leq$2                  & 170$\pm$50             & 8  \\
                 & XMM     & 3.82          & $<$3                    &   106                  &10  \\
NGC 4594         & XMM     & 1.3           & 18                      & $<$296                 &11  \\
NGC 4565         & Chandra & 0.22          &    25$\pm$6             &   --                   &12  \\
                 & XMM     & 0.24          &    12$\pm$4             &   --                   & 8   \\
NGC 4698         & Chandra & 0.013         &    5$^{+0.7}_{-0.5}$     &    --                  &13  \\ 
                 & XMM     & 0.04          & $<$40                   &    --                  &8   \\                 
                 & ASCA    & 1.04          &    9.5$^{+3.6}_{-4.2}$   & $<$425                 &14   \\
NGC 5033         & XMM     & 2.87          &$\leq$3                  & 466$\pm$215            &8   \\
                 & ASCA    & 2.6           & 0.27$^{+7.7}_{-0.27}$    &   --                   &15  \\                 
H1320+551        & XMM     & 2.1           & 1.69$^{+0.45}_{-0.39}$   & 380$^{+230}_{-320}$     &16    \\
Mrk 273x         & Chandra & 0.1           & 14.1$^{+5.5}_{-5.0}$     & $<$30                  &17   \\
NGC 5995         & ASCA    & 28.9          & 90$^{+5}_{-3}$           & 144$^{41}_{-41}$        &3   \\
NGC 6251         & Chandra & 4.5           & $<$3                    & --                     &18 \\     
                 & XMM     & 4.0           & 5.1$^{+1.1}_{-0.9}$      & 223$^{+219}_{-99}$      &19   \\ 
                 & ASCA    & 1.4           & 75$^{+64}_{-58}$         & 443$^{+313}_{-272}$     &20   \\      
IRAS 20051-1117  & Chandra & 1.5$\pm$0.02  & 0.1                     & 250$\pm$155            &21  \\
                 & XMM     & 1.5$\pm$0.025 & 3$^{+2}_{-1}$            & 360$^{+174}_{-170}$     &21  \\ 
                 & ASCA    & 2.4           & $<$40                   & 272$^{52}_{-73}$        & 3  \\                 
NGC 7590         & ASCA    & 1.2           & $<$9.2                  & --                     &22  \\
NGC 7679         & ASCA    & 5.2           & $<$7.6                  & --                     &23  \\
                 & SAX     & 6.0           & 2.2$^{+1.8}_{-1.4}$      & $<$ 180                &23  \\                 
RXJ1715.4+6239   & XMM     & 0.10          & $<$9                    & $<$ 250                &24  \\
RXJ1724.9+6636   & XMM     & 0.11          & 3-12                    & $<$ 190                &24  \\
RXJ1737.0+6601   & XMM     & 0.34          & $<$2                    & $<$240                 &24  \\
XBSJ031146.1-550702 & XMM  & 0.28          & $<$1.3                  & --                     &25  \\
$[$H2004$]$213004.24-430744.2 & XMM & 0.15    & $<$2                 & --                     &26  \\
$[$H2000$]$213115.90-424318.9 & XMM & 0.022   & $<$9                 & --                     &26  \\
\enddata
\tablecomments{Col.(1): Sources; Col.(2): The observatory for X-ray data; Col.(3): Observed 2-10 keV fluxes;
Col.(4): The HI column density. Col.(5): EW of Fe K$\alpha$ line. Col(6): References for X-ray data. Note that 
the confidence levels for errors and upper-limits are generally between 68\% and 90\%.\\
References: 1--\citet{Georgantopoulos03a}; 2--\citet{Georgantopoulos00}; 3--\citet{Panessa02}; 4--\citet{Corral05}; 
  5--\citet{Terashima03}; 6--\citet{Bianchi08a}; 7--\citet{Brightman08}; 8--\citet{Cappi06}; 9--\citet{Eracleous02};
 10--\citet{Dewangan04}; 11--\citet{Pellegrini03}; 12--\citet{Chiaberge06}; 13--\citet{Georgantopoulos03b};  
 14--\citet{Pappa01}; 15--\citet{Shinozaki06}; 16--\citet{Barcons03}; 17--\citet{Xia02}; 18--\citet{Evans05};           
 19--\citet{Gliozzi04}; 20--\citet{Sambruna99}; 21--\citet{Georgantopoulos04}; 22--\citet{Bassani99};     
 23--\citet{DellaCeca01}; 24--\citet{Wolter05}; 25--\citet{Caccianiga04}; 26--\citet{Panessa09}}

\end{deluxetable}

\begin{deluxetable}{lllcccclllll}
\tablecolumns{15}
\tabletypesize{\scriptsize}
\tablewidth{0pc}
\tablecaption{\label{sample_IR} Infrared Properties of X-ray-Unobscured Type 2 AGN Candidates}
\tablehead{
\colhead{sources}                             &
\colhead{IRS MODE}                            &
\colhead{$f_{12{\mu}m}^{\rm NUC}$}              &     
\colhead{$f_{24{\mu}m}^{\rm NUC}$}
\\
\colhead{}                                    &
\colhead{}                                    &
\colhead{[mJy]}             &
\colhead{[mJy]}       
\\
\colhead{(1)}                                 &
\colhead{(2)}                                 &
\colhead{(3)}                                 &
\colhead{(4)}                                 
}
\startdata

IRAS 00317-2142     &   STARE   &  --   &   --  \\
IRAS 01428-0404     &   --     &  --   &   --  \\
Mrk 993             &   STARE   &   11  &  41.1 \\
NGC 3147            &   STARE   & 30.5  &  84.0 \\
NGC 3660            &   MAP    &  1.1  &  33.2 \\
NGC 3941            &   STARE   &  2.2  &  14.5 \\
NGC 4565            &   STARE   & 17.8  &  47.9 \\
NGC 4579            &   MAP    & 14.8  &  69.5 \\
NGC 4594            &   MAP    &  2.2  &  24.9 \\
NGC 4698            &   STARE   &  1.7  &   4.6 \\
NGC 5033            &   MAP    &  --   &  33.8 \\
H1320+551           &   --     &  --   &   --  \\
Mrk 273x            &   --     &  --   &   --  \\
NGC 5995            &   MAP    &191.7  & 337.6 \\
NGC 6251            &   STARE   & 18.2  &  50.3 \\
IRAS 20051-1117     &   --     &  --   &   --  \\
NGC 7590            &   MAP    &  --   &  22.7 \\
NGC 7679            &   STARE   &  --   &   --  \\
RXJ 1715.4+6239     &   --     &  --   &   --  \\
RXJ 1724.9+6636     &   --     &  --   &   --  \\
RXJ 1737.0+6601     &   STARE   &  2.0  &   6.8 \\
XBSJ031146.1-550702 &   --     &  --   &   --  \\
$[$H2004$]$ 213004.24-430744.2 &  --   &   --  \\
$[$H2000$]$ 213115.90-424318.9 &  --   &   --  \\

\enddata
\tablecomments{Col.(1):   Sources.   Col.(2):   The  IRS   observation
mode. Col.(3): The nuclear flux density within IRAS 12 $\mu$m filter
after   subtracting  the   stellar  atmospheric   and  star-formation
emission.  The  associated  error   is  dominated  by  that  of  the
star-formation  component  derived from  the  aromatic  flux and  is
estimated to be $\sim$50\%  \citep{Shi07}. Col.(4): The nuclear flux
density  within the MIPS 24  $\mu$m  filter.  The  uncertainty is  also
dominated by the aromatic-derived  star-formation component and is about
50\%.}
\end{deluxetable}



\begin{deluxetable}{lllcccclllll}
\tablecolumns{15}
\tabletypesize{\footnotesize}
\tablewidth{0pc}
\tablecaption{\label{mis_classification} Summary of Mis-Classified X-ray-Unobscured Type 2 AGNs}
\tablehead{
\colhead{ source   }                          & 
\colhead{ Type 1   }                          & 
\multicolumn{3}{c}{ Compton-thick}           \\
 \cline{3-5}   \\
\colhead{}  & 
\colhead{}  & 
\colhead{$\frac{F_{\rm x}}{F_{\rm [OIIII]}}<$1} &
\colhead{EW(FeK$\alpha$)$>$1 KeV}             & 
\colhead{$L_{X}{\textendash}L_{\rm MIR}$}      \\
\colhead{(1)}  & 
\colhead{(2)}  & 
\colhead{(3)}  & 
\colhead{(4)}  & 
\colhead{(5)}    
}
\startdata
IRAS 00317-2142      &       Y                  &      &        &          \\
Mrk 993              &       Y                  &      &        &          \\
NGC 3660             &       Y                  &      &        &          \\
NGC 3941             &       N                  & -    &    -   &   Y      \\
NGC 4565             &       Y                  & N    &    N   &   Y      \\
NGC 4579             &       Y                  & N    &    N   &   N      \\
NGC 4698             &       N                  & Y    &    N   &   Y      \\
NGC 5033             &       Y                  &      &        &          \\
NGC 6251             &       Y                  &      &        &          \\
H1320+551            &       Y                  & N    &    N   &   -      \\
Mrk 273x             &       N?                 & N    &    N   &   -      \\
NGC 5995             &       Y                  & N    &    N   &   N      \\
IRAS 20051-1117      &       Y                  &      &        &          \\
NGC 7679             &       Y                  &      &        &          \\
RXJ 1715.4+6239      &       ?                  &      &        &          \\
RXJ 1724.9+6636      &       ?                  &      &        &          \\
RXJ 1737.0+6601      &       Y                  &      &        &          \\
XBSJ031146.1-550702  &       Y                  & -    &    -   &   -      \\
$[$H2004$]$ 213004.24-430744.2 & Y              &      &        &          \\
\enddata                      
\tablecomments{Col.(1): Sources. Col.(2): $'$Y$'$ indicates the type 1 classification 
while $'$N$'$ is for type 2. $'$?$'$ indicates low S/N for the optical spectra.
Col.(3)-Col.(5):  Compton-thick objects with solid type 2 classifications are
marked as $'$Y$'$. A $'$-$'$ is for the cases with no available data.
}

\end{deluxetable}

\clearpage
\begin{deluxetable}{llllllllllllllll}
\tabletypesize{\scriptsize}
\tablewidth{0pc}
\tablecaption{\label{SAMPLE_TRUE} Properties of Possible X-ray-Unobscured Type 2 AGNs}
\tablehead{     
\colhead{sources}           &
\colhead{Type}              &
\colhead{Chandra}           &
\colhead{SIMU.}             &
\colhead{10-20${\mu}$m}     &
\colhead{PBEL}              &
\colhead{Ref.}              &
\colhead{Log$L_{\rm BOL}$}   &
\colhead{Log$(M_{\rm BH})$}  &
\colhead{Ref}               &
\colhead{V$_{\rm X{\textendash}ray}$} &
\colhead{Ref}                       &
\colhead{V$_{\rm UV}$}               &
\colhead{Ref}                       
 \\
\colhead{(1)}            &
\colhead{(2)}            &
\colhead{(3)}            &
\colhead{(4)}            &
\colhead{(5)}            &
\colhead{(6)}            &
\colhead{(7)}            &
\colhead{(8)}            &
\colhead{(9)}            &
\colhead{(10)}           &
\colhead{(11)}           &
\colhead{(12)}           &
\colhead{(13)}           &
\colhead{(14)}           
} 
\startdata
IRAS 01428-0404                & S2                      &     &           &  --       &     &     &  41.99    &  7.3    [$M_{\rm Bulge}$]    &  3  &     &     &    &    \\
NGC 3147                       & S2                      & Y   &  Y        &Aromatic   & N   &  1  &  43.29    &  8.4    [$M_{\rm Bulge}$]    &  4  &     &     &    &    \\
NGC 4594                       & L2                      &     &           &Sil. Emi.  &     &     &  41.01    &  8.3    [$M_{\rm Bulge}$]    &  6  & N   &  10 & Y  &13 \\
NGC 7590                       & S2                      &     &           &Aromatic   & N   &  2  &  41.19    &  6.79    [$\sigma_{*}$]     &  8  &     &     &    &   \\ 
$[$H2000$]$ 213115.90-424318.9 & S2          &     &   Y       &  --       &     &     &           &                             &     &     &     &    &   \\

\enddata
\tablecomments{
Col.(1): Sources.  Col.(2): Optical  AGN type.  Col.(3): 'Y' indicates
that there  is Chandra observation.  Col.(4): 'Y'  indicates the X-ray
and  optical  observation  are  taken  simultaneously.   Col.(5):  the
spectral feature  at 10-20 $\mu$m. Col.(6): The  presence of polarized
broad  emission line  (PBEL). Col.(7):  The references  for  the PBEL.
Col.(8):  The bolometric  luminosity  in erg  s$^{-1}$.  Col.(9):  The
black hole  mass in M$_{\odot}$  corrected to the distance  adopted in
this  paper.  '$M_{\rm  Bulge}$' indicates  the BH  mass based  on the  
bulge mass.  '$\sigma_{*}$' indicates the BH mass based on the stellar
velocity dispersion. 'H$\alpha$' indicates  the BH mass using the
H$\alpha$ emission  line.   'DYN.' indicates  the BH  mass
calculated through a dynamical model.  Col.(10): The references for BH
masses.  Col.(11): The variability  in X-ray. Col.(12): References for
X-ray  variability.   Col.(13):   The  variability  in  UV.  Col.(14):
References for UV variability. \\
References:  1 - This work;  2 - \citet{Heisler97}; 3 - \citet{Wang07}; 
 4 - \citet{Dong06}; 5 - \citet{Wang07}; 6 - \citet{Dong06};  7 - \citet{Ferrarese99}; 
 8 - \citet{Bian07}; 9 -- \citet{Eracleous02}; 10- \citet{Pellegrini03}; 11- \citet{Gliozzi04};
 12-- \citet{Maoz05}; 13-- \citet{Maoz05}; 14-- \citet{Evans05}
}
\end{deluxetable}
\clearpage


\begin{deluxetable}{llllllllllllll}
\tabletypesize{\scriptsize}
\tablewidth{0pc}
\tablecaption{\label{NSED_3147} Nuclear SED of NGC 3147 }
\tablehead{     
\colhead{Freq.}           & \colhead{Flux}                          &    \colhead{Ref.}        \\
\colhead{[Hz]}           &  \colhead{[erg s$^{-1}$ cm$^{-2}$]}         } 
\startdata
  1.63E+09 &  9.78E-17 &     \citet{Krips07} \\
  1.70E+09 &  1.14E-16 &  \citet{Anderson04} \\
  2.30E+09 &  1.50E-16 &  \citet{Anderson04} \\
  4.99E+09 &  4.64E-16 &     \citet{Krips07} \\
  5.00E+09 &  4.20E-16 &  \citet{Anderson04} \\
  8.40E+09 &  7.48E-16 &  \citet{Anderson04} \\
  1.54E+10 &  1.25E-15 &  \citet{Anderson04} \\
  2.22E+10 &  1.86E-15 &  \citet{Anderson04} \\
  4.32E+10 &  4.54E-15 &  \citet{Anderson04} \\
  9.99E+10 &  5.30E-15 &     \citet{Krips06} \\
  3.00E+11 &  8.40E-15 &     \citet{Krips06} \\
  1.25E+13 &  1.69E-11 &           This-work \\
  2.50E+13 &  8.89E-12 &           This-work \\
  1.45E+18 &  3.70E-12 & \citet{Terashima03} \\
\enddata
\end{deluxetable}

\begin{deluxetable}{llllllllllllll}
\tabletypesize{\scriptsize}
\tablewidth{0pc}
\tablecaption{\label{NSED_4594} Nuclear SED of NGC 4594 }
\tablehead{     
\colhead{Freq.}           & \colhead{Flux}                          &    \colhead{Ref.}        \\
\colhead{[Hz]}           &  \colhead{[erg s$^{-1}$ cm$^{-2}$]}         } 
\startdata
  1.50E+10 &  1.50E-14 &    \citet{Hummel84} \\
  1.25E+13 &  3.00E-12 &           This-work \\
  2.50E+13 &  3.24E-13 &           This-work \\
  9.09E+14 &  5.05E-13 &      \citet{Maoz05} \\
  1.20E+15 &  1.88E-13 &      \citet{Maoz05} \\
  3.02E+17 &  5.50E-13 &\citet{Pellegrini03} \\
  1.45E+18 &  1.30E-12 &\citet{Pellegrini03} \\
\enddata
\end{deluxetable}


\clearpage

\end{document}